# Strontium Titanate Capping Layers as Kinetic Regulators of Redox Phase Transitions in Oxide Heterostructures

Pooya Abdollahi[1], Woo Jin Kim[2,3,4], Keaghan Knight[1], Harold Y. Hwang[3,4], Ambrose Seo[1,*]

[1] *Department of Physics and Astronomy, University of Kentucky, Lexington, KY 40506, USA*

[2] *School of Materials Science and Engineering, Pusan National University, 2, Busandaehak-ro 63-beon-gil, Busan, Geumjeong-gu, 46241, Republic of Korea*

[3] *Department of Applied Physics, Stanford University, CA 94305, USA*

[4] *Stanford Institute for Materials and Energy Sciences, SLAC National Accelerator Laboratory, Menlo Park, CA 94025, USA*

**Abstract:**

Oxide capping layers are widely used to protect and stabilize thin-film heterostructures, yet their role in governing phase-transition kinetics remains largely unexplored. Here, we uncover the kinetic function of few-nanometer-thick $SrTiO_3$ (STO) capping layers in controlling oxidation and reduction processes, using *real-time in situ* optical spectroscopic ellipsometry. We show that oxygen diffusion through the nanoscale capping layer constitutes the rate-limiting step, giving rise to diffusion-controlled phase transitions in the underlying materials. As a result, both capping-layer thickness and oxygen partial pressure provide effective handles to tune phase transition temperatures. Quantitative analysis yields an activation energy of $E_A = 0.64 \pm 0.06$ eV for oxygen diffusion through STO, highlighting its suitability for precise control of redox reactions in diverse material systems. These results establish oxide capping layers as active kinetic regulators of redox-driven phase transitions and offer a general framework for engineering redox dynamics in complex oxide heterostructures.

* E-mail: a.seo@uky.edu

## 1. Introduction

Ultrathin solid-state capping layers or encapsulation layers are widely employed in the design of advanced materials and heterostructures. Beyond their conventional role in surface passivation and protection,[1-11] capping layers can impact diverse functionalities. For example, Pd-Au alloy caps enhance hydrogenation kinetics in metal hydride-based systems such as hydrogen sensors,[12] while nanometer-thin $CeO_2$ layers facilitate Li-ion intercalation at electrolyte-electrode interfaces in lithium-ion batteries.[13] Capping layers have also been used to enhance the quantum anomalous Hall effect with $CrO_x$ caps,[14] improve field-effect mobility in p-channel SnO transistors with Ni caps,[15] and modify the electronic response at $LaAlO_3/SrTiO_3$ interfaces with Ti caps.[16]

Among candidate materials, strontium titanate, $SrTiO_3$ (STO), has emerged as a prototypical capping layer in complex oxide systems. STO caps not only enable controlled reduction by extracting oxygen from thin films[17-19] but can also drive structural transitions, such as conversion of tetragonal $SrRuO_3$ into the orthorhombic phase.[20] More notably, STO plays a vital role in stabilizing metastable phases under nonequilibrium conditions, enabling the realization of superconducting nickelates[21-23] and magnetic cobaltates.[24]

While capping layers are widely recognized for their roles in protection, stabilization, and property tuning, their influence on the kinetics of phase transitions in underlying films remains poorly understood. This is particularly important because even subtle variations in growth rate[25,26] and processing conditions such as temperature[27,28] and oxygen pressure[29,30] can dramatically alter thin-film structure and quality. Although capping layers are expected to modify these processes, their role in governing phase-transition kinetics has largely remained empirical and lacks a quantitative framework.

In this article, we directly address this question by investigating how capping layers control the phase-transition kinetics of oxide thin films. Using *real-time in situ* optical spectroscopic ellipsometry, we track the oxidation and reduction dynamics of $CaCoO_x$ ($x$ = 2.0 or 2.5) thin films capped with ultrathin STO layers. By systematically varying capping-layer thickness, oxygen partial pressure, and temperature, we demonstrate that the phase transition follows a diffusion-controlled mechanism, with oxygen transport through the capping layer as the rate-limiting step. The extracted activation energies for oxygen diffusion are consistent with reported values for STO and enable quantitative prediction of transition temperatures under varying conditions. These results establish that capping layers are not merely passive protective coatings, but they function as diffusion-controlled kinetic gates that quantitatively govern topotactic phase transitions in complex oxides.

## 2. Results

### 2.1. Experimental Approach and *In Situ* Spectroscopic Ellipsometry Framework

To probe the effects of capping layers, we focus on metastable $CaCoO_2$ thin films capped with a few unit cells (u.c.) of STO. Brownmillerite $CaCoO_{2.5}$ films are first grown by pulsed laser deposition on STO (001) substrates, followed by topotactic reduction to form the infinite-layer $CaCoO_2$ phase.[24,31] The $CaCoO_2$ thin films used in this study have the same thickness of 18 ± 2 nm, confirmed by the X-ray diffraction (XRD) and X-ray reflectivity (XRR) scans (Supplementary Figure S1a,b) and spectroscopic ellipsometry measurements. Because this infinite-layer structure is metastable, it readily transforms back to the brownmillerite phase upon oxidation, as schematically illustrated in Figure 1a. This phase transition is accompanied by pronounced changes in the optical response, making the STO/$CaCoO_x$ system suited for real-time detection of oxygen transport and phase evolution through capping layers.

Characterizing thin-film heterostructures is inherently challenging because the layers of interest are often buried beneath capping layers, limiting direct access. *Real-time* measurements further require techniques that do not rely on separate reference samples. To address these challenges, we employ *in-situ* optical spectroscopic ellipsometry integrated with a vacuum chamber. As illustrated in Figure 1b, spectroscopic ellipsometry probes the change in polarization of light upon reflection, described by the ellipsometric angles, $\Psi$ and $\Delta$.[32] The measured complex reflectance ratio is given by $\tilde{\rho}(E) = \frac{\tilde{r}_p(E)}{\tilde{r}_s(E)} = \tan\Psi(E)e^{i\Delta(E)}$, where $\tilde{r_s}$ and $\tilde{r_p}$ are the Fresnel reflection coefficients of *s*- and *p*-polarized light, respectively. Here, $\tan\Psi$ represents the amplitude ratio, while $\Delta$ corresponds to the phase difference between the two polarization components. From $\Psi$ and $\Delta$, one can directly extract key optical and structural parameters, including film thickness and complex dielectric functions.[33,34] The combination of *in-situ* operation, *real-time* monitoring,[35,36] long probing depth,[37] and self-normalizing nature[38] makes spectroscopic ellipsometry a powerful and versatile tool for characterizing buried layers and evolving material systems.

### 2.2. Diffusion Controlled Kinetics by Ultrathin $SrTiO_3$ Capping Layers

Our experimental results demonstrate that the presence of capping layers directly modifies the phase-transition behavior of the underlying thin film. To investigate this effect, $CaCoO_{2+\delta}$ ($0 \leq \delta < 0.5$) films capped with 6 u.c. STO were subjected to isothermal oxidation at different temperatures under 500 Torr of oxygen partial pressure ($P_{O2}$). Although these films do not exhibit a perfectly stoichiometric infinite-layer phase, they serve as a sensitive probe of oxygen incorporation. The evolution of the ellipsometry spectra (Ψ and Δ) during oxidation is shown in Supplementary Figure S2a-e for five temperatures (170 °C, 205 °C, 230 °C, 245 °C, and 275 °C), recorded at one-minute intervals. The most pronounced spectral changes occur near 3.9 eV. Accordingly, Figure 2a presents the time evolution of $\Psi(t)$ at 3.9 eV for these temperatures. Notably, all temperatures lie below the phase-transition temperature required for capped $CaCoO_2$, as shown in Figure 3b, and the optical response evolves gradually under these conditions.

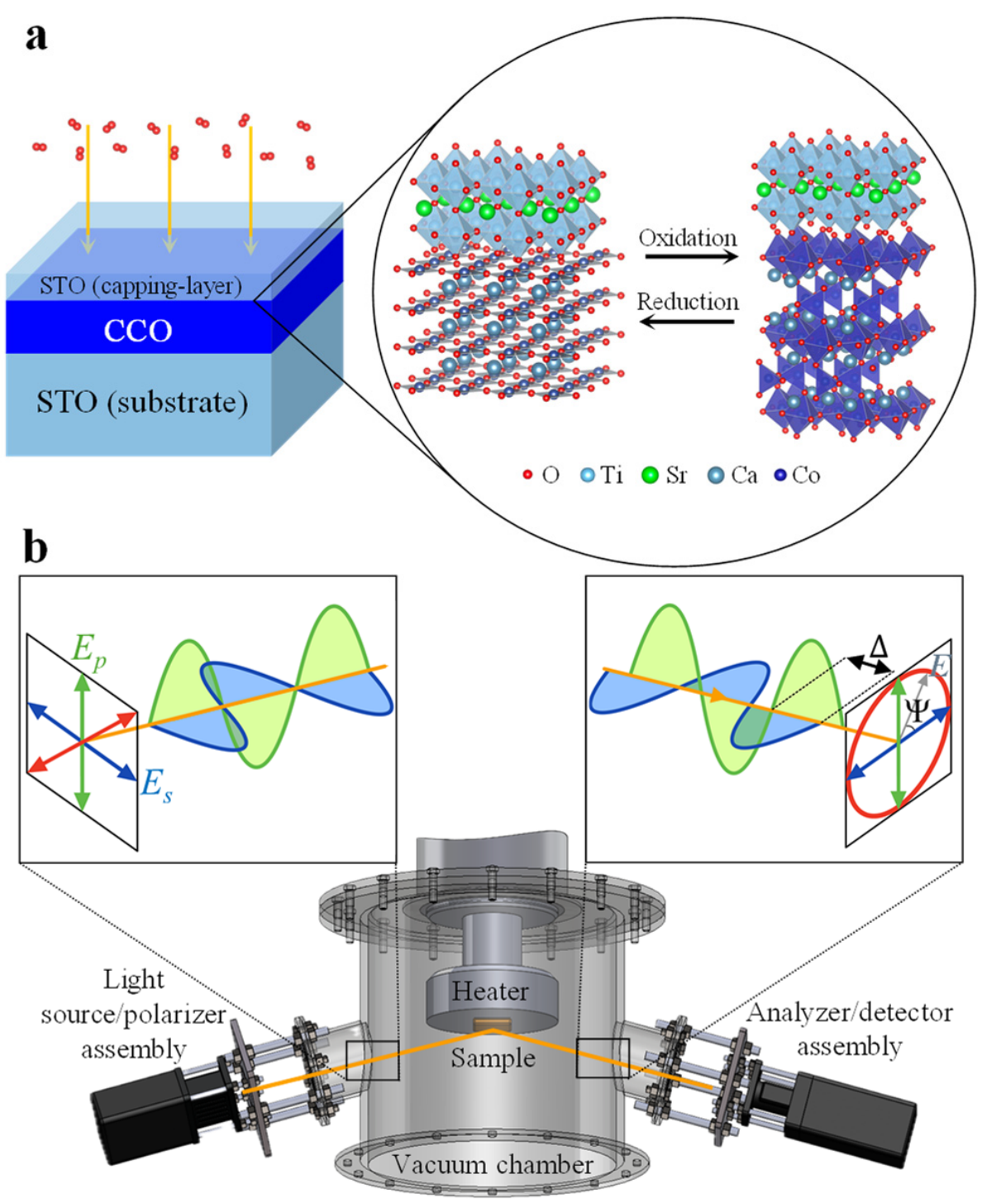

**Figure 1.** (a) Schematic diagram of the studied sample structure, showing oxygen diffusion through ultrathin $SrTiO_3$ (STO) capping layers toward $CaCoO_x$ ($x$ = 2.0-2.5) films deposited on STO substrates. The inset illustrates the reversible topotactic phase transition between the infinite-layer $CaCoO_2$ structure and the brownmillerite $CaCoO_{2.5}$ structure upon oxidation and reduction, respectively. (b) Schematic diagram of the experimental setup consisting of a vacuum chamber equipped with *in situ* optical spectroscopic ellipsometry.

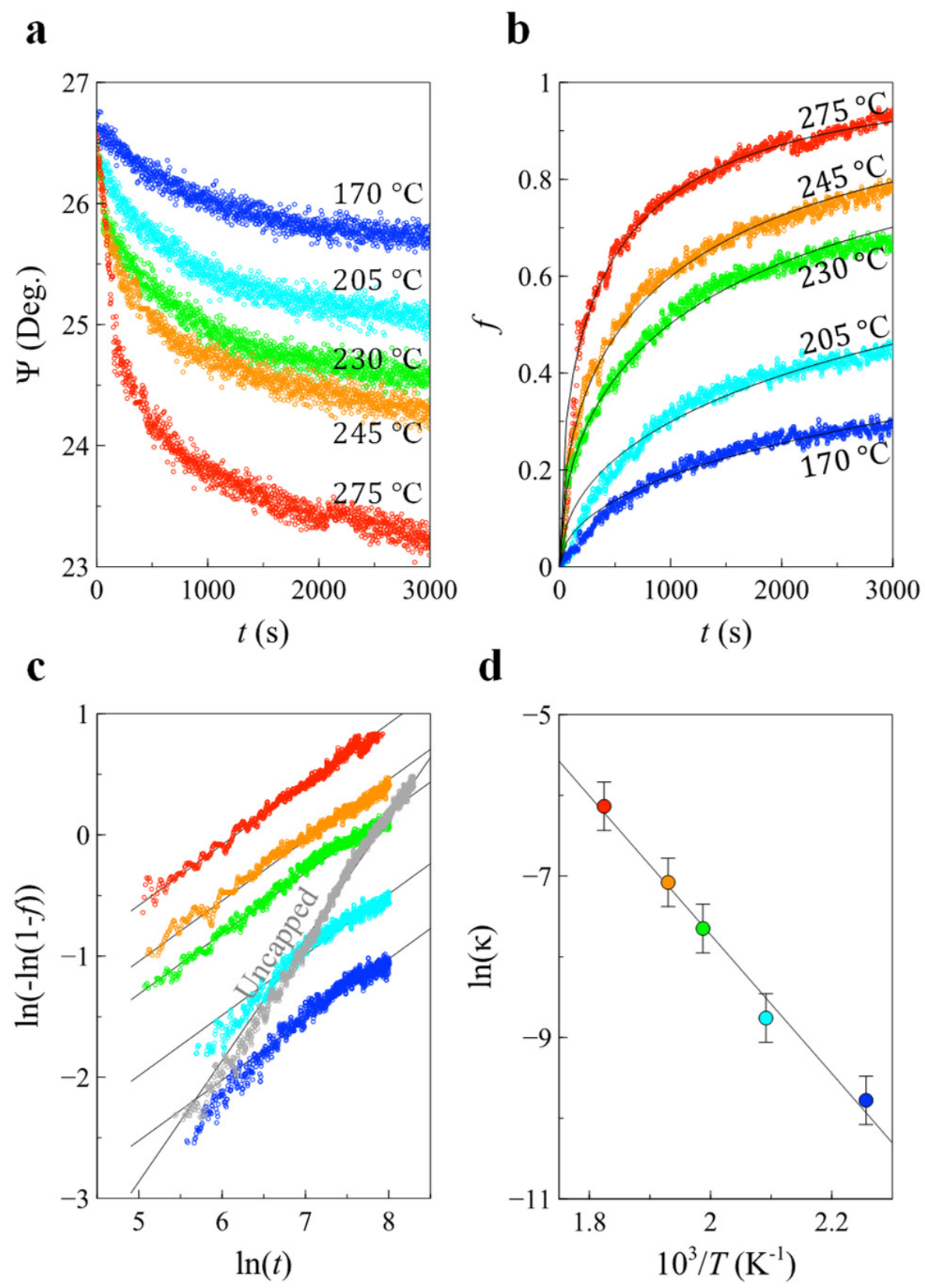

**Figure 2.** Oxidation of $CaCoO_2$ films with 6 u.c. STO capping layers under various temperatures. (a) Time evolution of Ψ at 3.9 eV during oxidation at five different temperatures. (b) Volume fraction of oxidation, *f*, as a function of time at the corresponding isothermal conditions. (c) Avrami analysis of the oxidation kinetics. The solid lines show Avrami exponents, $n = 0.5$ for STO-capped films (the same colors as in (b)) and $n = 1$ for an uncapped $SrCoO_x$ film (grey color). (d) Arrhenius plot of the rate constant, showing ln(κ) as a function of the inverse temperature ($10^3/T$). The slope indicates the activation energy of oxygen diffusion (~0.7 eV) in STO capping layers.

To quantify the transition kinetics, we employ the Johnson-Mehl-Avrami-Kolmogorov (JMAK) model,[39,40] which describes the phase transition fraction as $f(t) = 1 - e^{-at^n}$. Here, $f(t)$ represents the fraction of oxidation after time $t$. We estimate the fraction of oxidation from the ellipsometry data via $f_{3.9eV}(t) \equiv [\Psi_{3.9\,\text{eV}}(t) - \Psi_{3.9\,\text{eV}}(t=0)]/[\Psi_{3.9\,\text{eV}}(t=\infty) - \Psi_{3.9\,\text{eV}}(t=0)]$, as shown in Figure 2b. In the JMAK equation, $a$ and $n$ are the Avrami coefficients corresponding to the rate constant and time exponent, respectively, and provide insight into the phase transition mechanism. The Avrami exponent can be expressed as $n = \frac{d}{m} + b$, where $d$ is the dimensionality of phase transition, $m$ distinguishes interface-controlled ($m$ = 1) from diffusion-controlled ($m$ = 2) kinetics, and $b$ characterizes nucleation behavior. For heterogeneous nucleation, the number of nuclei is time-independent ($b$ = 0), whereas homogeneous nucleation yields $b$ = 1.[41-43]

From fits to the data (Figure 2c), we obtain $n$ = 1/2, indicating a heterogeneous ($b$ = 0), one-dimensional ($d$ = 1), and diffusion-limited ($m$ = 2) phase transition. This behavior is consistent with oxygen diffusion through the STO capping layer as the rate-limiting step. After traversing the capping layer, oxygen nucleates an oxidized phase at the STO/$CaCoO_{2+\delta}$ interface, which then propagates unidirectionally into the film. For comparison, we performed parallel measurements on uncapped $SrCoO_{2+\delta}$ ($0 \leq \delta < 0.5$) films. $SrCoO_{2+\delta}$ is used because uncapped $CaCoO_{2+\delta}$ films are unstable, while both systems share similar cobaltate frameworks and topotactic oxidation behavior. As shown in Figure 2c, the uncapped films exhibit $n$ = 1, characteristic of interface-controlled kinetics ($m$ = 1). This contrast demonstrates that the capping layer fundamentally alters the phase transition process, shifting it from interface-controlled to diffusion-controlled behavior.

The Avrami parameters are related to the Arrhenius rate constant via $\kappa = a^{\frac{1}{n}}$ [Refs.[44-47]]. By fitting the temperature dependence of κ using $\kappa(T) = \kappa_0\, e^{-\frac{E_A}{k_B T}}$, as shown in Figure 2d, we extract an activation energy $E_A$ = 0.7 ± 0.1 eV, consistent with reported values of oxygen diffusion in STO (0.5 - 0.9 eV).[48-51]

### 2.3. Capping Layer Thickness-Dependent Effects on Redox Phase Transition Temperatures

Having observed that the capping layer governs the redox kinetics, we next examine the effect of capping layer thickness. Because parameters such as temperature can influence film structure and quality,[27,28] it is important to isolate capping layer thickness-dependent effects. Infinite-layer $CaCoO_2$ films capped with 2 and 6 u.c. STO were subjected to oxidation under $PO_2$ = 500 Torr. Figure 3a shows the evolution of Ψ and Δ for the 2 u.c. sample (data for 6 u.c. shown in Supplementary Figure S3a,b). To facilitate comparison, we track Ψ and Δ at four representative photon energies (Figure 3b). Increasing the capping layer thickness from 2 u.c. to 6 u.c. shifts the spectral changes to higher temperature, indicating an increase in the phase transition temperature from $T_{2u.c.}$ = 511 ± 2 K to $T_{6u.c.}$ = 553 ± 2 K. This observation further supports a diffusion-limited mechanism, where oxygen transport through the capping layer controls the phase transition kinetics.

A similar thickness dependence is observed during reduction. Figure 3c and 3d show the evolution of Ψ and Δ for $CaCoO_{2.5}$ during reduction under vacuum (< $10^{-7}$ Torr). Although the reduction transition is less abrupt than oxidation, clear differences in transition temperatures between the 2 and 6 u.c. samples reinforce the role of the capping layer in controlling redox kinetics.

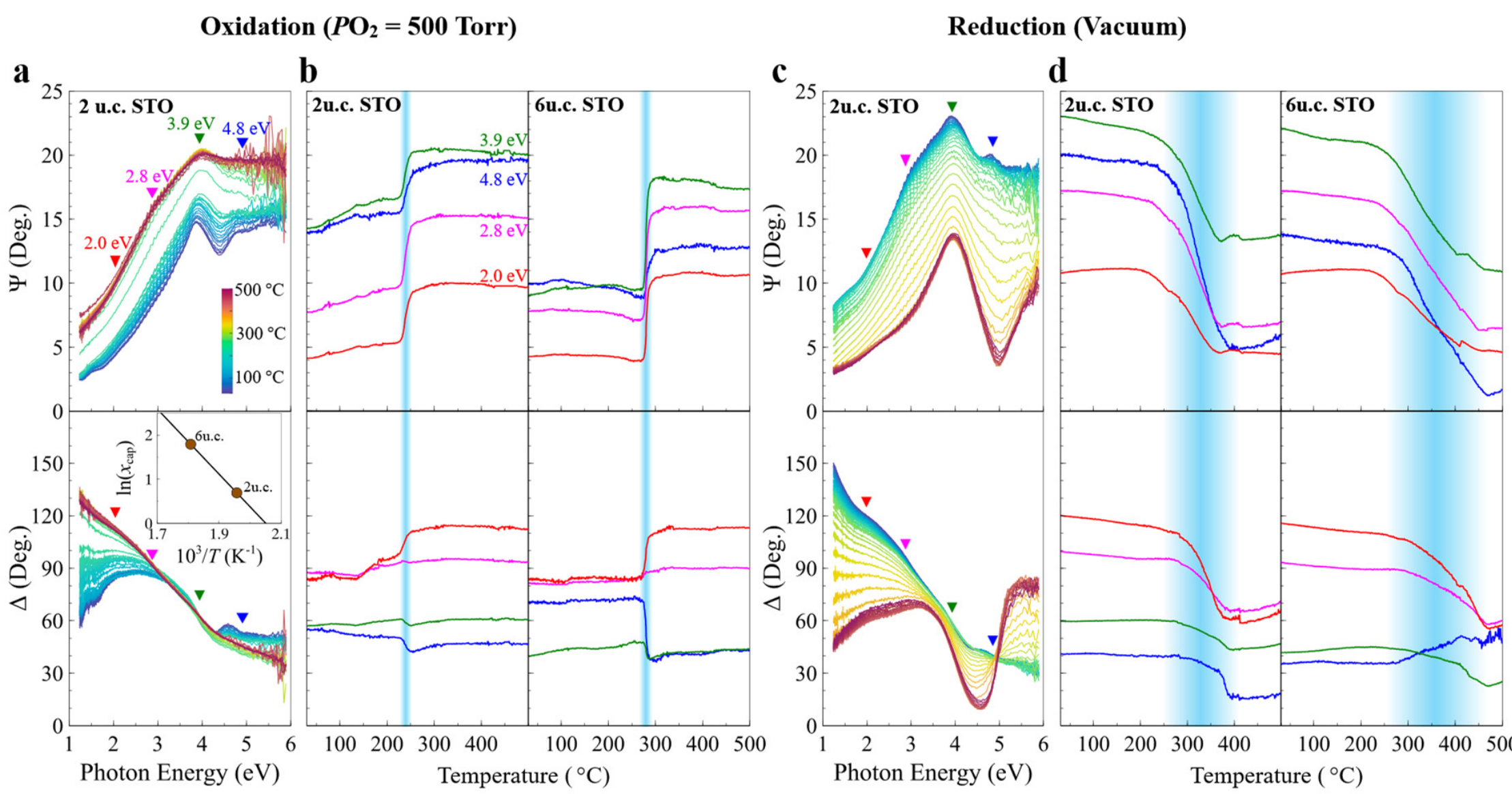

**Figure 3.** Spectroscopic ellipsometry data during oxidation and reduction of STO-capped $CaCoO_x$ films. (a) Ellipsometry parameters Ψ and Δ as a function of photon energies during the oxidation of $CaCoO_2$ with 2 u.c. STO capping layer under $PO_2$ = 500 Torr. Spectra are shown at temperature intervals of 10 °C. The inset shows the logarithm of capping layer thickness ($\ln(x_{cap})$) vs. the inverse phase transition temperatures ($10^3/T$). The slope gives the activation energy of ~0.64 eV. (b) Temperature dependence of Ψ and Δ at four selected photon energies (2.0, 2.8, 3.9, and 4.8 eV) during oxidation. Phase transition temperature (vertical blue shades) increases from $T_{2u.c.}$ = 238 °C (511 K) to $T_{6u.c.}$ = 280 °C (553 K) as increasing STO capping layers. (c) Ellipsometry parameters Ψ and Δ as a function of photon energies during the reduction of $CaCoO_{2.5}$ with 2 u.c. STO capping layer under vacuum (< $10^{-7}$ Torr). Spectra are shown at temperature intervals of 10 °C. (d) Temperature dependence of Ψ and Δ at four selected photon energies during reduction. Different phase transition temperatures (vertical blue shades) are observed between samples with 2 u.c. (≈ 320 °C) and 6 u.c. (≈ 360 °C) STO capping layers, although the transitions are less abrupt than during oxidation.

These results are quantitatively consistent with a one-dimensional diffusion model. According to Fick's law,[52] the diffusion flux is described by $J = -D\,\frac{dc}{dx}$, where $D$ is the diffusion coefficient, $c$ is the oxygen concentration, and $x$ is the thickness (capping layer). Given that a minimum threshold of oxygen flux $J$ is necessary to induce the phase transition, the thicker (6 u.c.) STO capping layer requires a proportionally larger diffusion coefficient than the thinner (2 u.c.) layer. Using the Arrhenius form $D(T) = D_0\,e^{-\frac{E_A}{k_B T}}$ and the measured oxidation temperatures ($T_{2u.c.} = 511 \pm 2$ K, $T_{6u.c.} = 553 \pm 2$ K), we extract $E_A = 0.64 \pm 0.06$ eV, in good agreement with the independent estimate above.

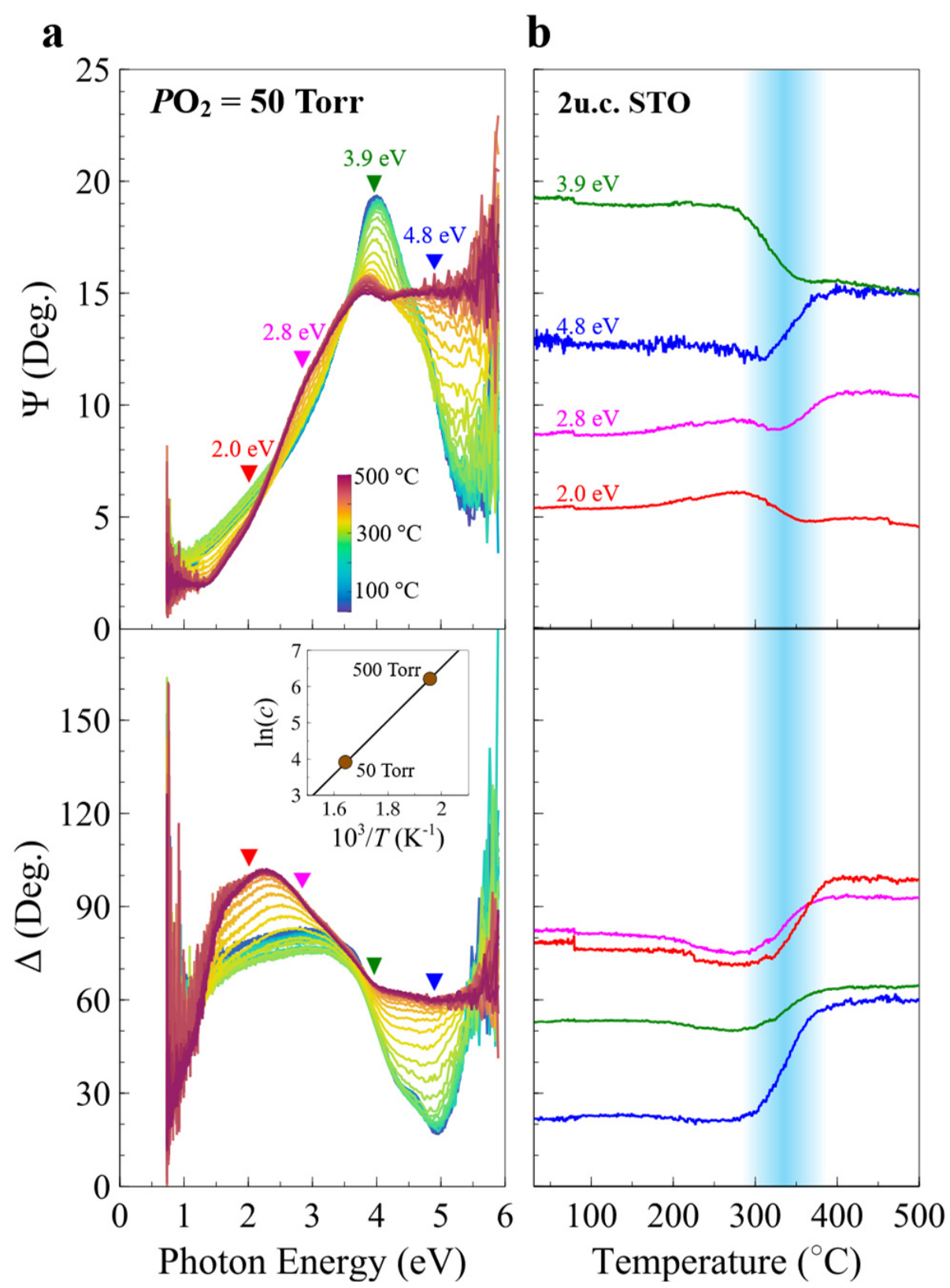

**Figure 4.** Spectroscopic ellipsometry data during the oxidation of $CaCoO_2$ with 2 u.c. STO capping layer at $PO_2 = 50$ Torr. (a) Ellipsometry parameters Ψ and Δ as a function of photon energies. Spectra are shown at temperature intervals of 10 °C. The inset shows the logarithm of oxygen pressure (ln($c$)) vs. inverse phase transition temperatures ($10^3/T$). The slope shows the activation energy of ~0.64 eV. (b) Temperature dependence of Ψ and Δ at four selected photon energies during the oxidation process, showing a phase transition temperature (vertical blue shades) $T_{50Torr} \approx 340$ °C ($PO_2 = 50$ Torr), which is increased by ~100 °C from that of $PO_2 = 500$ Torr.

### 2.4. Oxygen Pressure Effects on Redox Phase Transition Temperatures

Finally, we examine the effect of oxygen pressure. With the capping layer thickness fixed at 2 u.c., oxidation was performed at $PO_2$ = 50 Torr, which is an order of magnitude lower pressure than 500 Torr, as shown in Figure 4a. With the same assumption that a critical oxygen flux is required to induce the topotactic phase transition, the diffusion coefficient at $PO_2$ = 50 Torr must exceed that at 500 Torr. Using the extracted activation energy of ~0.64 eV, we can expect a new transition temperature of ~336 °C at $PO_2$ = 50 Torr. Our experimental data (Figure 4b) show excellent agreement with this value, $T_{50\mathrm{Torr}}$ = 340 ± 10 °C, further validating the oxygen diffusion model through the capping layer.

These results demonstrate that capping layers provide a powerful means of controlling phase-transition kinetics, with thickness, temperature, and oxygen pressure serving as tunable parameters that govern oxygen transport and, consequently, the transformation dynamics.

## 3. Conclusion

In conclusion, we demonstrate that ultrathin capping layers function as diffusion-controlled kinetic gates that quantitatively govern topotactic phase transitions in complex oxide thin films. Using *real-time*, *in-situ* spectroscopic ellipsometry, we identify oxygen diffusion through the STO capping layer as the rate-limiting step, leading to a diffusion-controlled phase transition. Systematic variation of capping-layer thickness, oxygen partial pressures, and temperature reveals that these parameters directly modulate the transition dynamics. The extracted activation energy ($E_A$ = 0.64 ± 0.06 eV) is consistent with reported values for oxygen diffusion in STO, validating our diffusion-based framework.

Beyond establishing the mechanistic role of capping layers, our results highlight their utility as a tunable handle for controlling phase-transition kinetics in buried oxide systems. While this study focuses on oxidation and reduction, the same approach is broadly applicable to other topotactic transitions involving mobile ionic species. In particular, extending this framework to systems such as lithium-containing oxides, where interfacial ion transport is critical,[53,54] offers a promising pathway for engineering reaction kinetics through interface design. More generally, our findings open new opportunities for leveraging capping layers to tailor ion transport, reaction pathways, and ultimately functional properties in thin-film heterostructures and devices.

## 4. Experimental Section/Methods

*Synthesis of $CaCoO_2$ thin films:* Brownmillerite $CaCoO_{2.5}$ thin films were grown by pulsed laser deposition on perovskite $SrTiO_3$ (001) substrates using a polycrystalline $CaCoO_{2.5}$ target and a krypton fluoride excimer laser (wavelength $\lambda = 248$ nm, repetition rate 3 Hz, fluence 1 J $cm^{-2}$). The target-to-substrate distance was fixed at 50 mm. Optimal film quality was achieved at a substrate temperature of $T = 600$ °C under an oxygen partial pressure of $P_{O2} = 50$ mTorr. An ultrathin $SrTiO_3$ capping layer was then deposited on the as-grown $CaCoO_{2.5}$ films to protect against degradation during subsequent reduction. Topotactic reduction to $CaCoO_2$ was carried out by sealing the capped films with approximately 0.1 g of $CaH_2$ powder in a vacuum glass tube, with the film loosely covered by aluminum foil. The sealed tube was annealed at 250 °C for 3 h in a tube furnace, with thermal ramp and cooling rates of 10 °C $min^{-1}$.

*Real-Time In Situ Spectroscopy Ellipsometry*: Optical spectroscopic ellipsometry was performed using a J.A. Woollam M-2000 ellipsometer over a photon energy range of 1.24–5.89 eV at an incident angle of 65°. The ellipsometer was attached to a vacuum chamber equipped with an infrared radiant heater, enabling *in situ* characterization during thermal cycling. Temperature was ramped at ~17 °C $min^{-1}$, and ellipsometry spectra were acquired every 2 s.

**Acknowledgements**

We thank Sujan Shrestha for assisting with the experiments. This work was supported by National Science Foundation Grant No. DMR-2426874 and the University of Kentucky Energy Research Priority Area. WJK acknowledges a Korea Basic Science Institute (National Research Facilities 3 and Equipment Center) grant funded by the Ministry of Science and ICT (No. RS-2024-400404737), and New Faculty Research and 2-5 Year Research grants of Pusan National University. Work at Stanford/SLAC was supported by the U.S. Department of Energy, Office of Basic Energy Sciences, Division of Materials Sciences and Engineering under Contract No. DE-AC02-76SF00515.

**Supporting Information**

Supporting Information, including XRD, XRR, spectroscopic ellipsometry spectra, and the method of determining phase transition temperature (Figures S1-S5), is available upon requests to AS (a.seo@uky.edu).

**References**

(1) X. Zhao, T. Liu, A.B. Kaplan, C. Yao, Y.L. Loo. Accessing Highly Oriented Two-Dimensional Perovskite Films via Solvent-Vapor Annealing for Efficient and Stable Solar Cells. *Nano Lett* **20**, 8880-8889 (2020). https://doi.org:10.1021/acs.nanolett.0c03914

(2) O. Djema, A. Moussi, S. Chaouchi, et al. Ag and Sn capping layers for stabilizing Cu solar cell contacts against oxidation in air: effects of method and solution of deposition. *Journal of Materials Science: Materials in Electronics* **35** (2024). https://doi.org:10.1007/s10854-023-11794-9

(3) J. Park, Y.A. Soh, G. Aeppli, et al. Crystallinity of tellurium capping and epitaxy of ferromagnetic topological insulator films on $SrTiO_3$ *Sci Rep* **5**, 11595 (2015). https://doi.org:10.1038/srep11595

(4) C.M. Sutter-Fella, Y. Li, M. Amani, et al. High Photoluminescence Quantum Yield in Band Gap Tunable Bromide Containing Mixed Halide Perovskites. *Nano Lett* **16**, 800-806 (2016). https://doi.org:10.1021/acs.nanolett.5b04884

(5) H. Lu, M. Rossi, A. Nag, et al. Magnetic excitations in infinite-layer nickelates. *Science* **373**, 213-216 (2021). https://doi.org:doi:10.1126/science.abd7726

(6) P. Salg, L. Zeinar, A. Radetinac, et al. Oxygen diffusion barriers for epitaxial thin-film heterostructures with highly conducting $SrMoO_3$ electrodes. *Journal of Applied Physics* **127** (2020). https://doi.org:10.1063/1.5129767

(7) I.S. Zhidkov, A.I. Kukharenko, M.A. Milyaev, et al. Protection of Cu from Oxidation by Ta Capping Layer. *Coatings* **13** (2023). https://doi.org:10.3390/coatings13050926

(8) L. Gan, R.D. Gomez, C.J. Powell, et al. Thin Al, Au, Cu, Ni, Fe, and Ta films as oxidation barriers for Co in air. *Journal of Applied Physics* **93**, 8731-8733 (2003). https://doi.org:10.1063/1.1543873

(9) D. Li, C. Adamo, B.Y. Wang, et al. Stabilization of $Sr_3Al_2O_6$ Growth Templates for Ex Situ Synthesis of Freestanding Crystalline Oxide Membranes. *Nano Lett* **21**, 4454-4460 (2021). https://doi.org:10.1021/acs.nanolett.1c01194

(10) P. Cao, P. Bai, A.A. Omrani, et al. Preventing Thin Film Dewetting via Graphene Capping. *Advanced Materials* **29** (2017). https://doi.org:10.1002/adma.201701536

(11) S. Sellner, A. Gerlach, F. Schreiber, et al. Strongly Enhanced Thermal Stability of Crystalline Organic Thin Films Induced by Aluminum Oxide Capping Layers. *Advanced Materials* **16**, 1750-1753 (2004). https://doi.org:10.1002/adma.200400461

(12) L.J. Bannenberg, B. Boshuizen, F.A. Ardy Nugroho, H. Schreuders. Hydrogenation Kinetics of Metal Hydride Catalytic Layers. *ACS Appl Mater Interfaces* **13**, 52530-52541 (2021). https://doi.org:10.1021/acsami.1c13240

(13) Y. He, H. Pham, X. Liang, J. Park. Impact of ultrathin coating layer on lithium-ion intercalation into particles for lithium-ion batteries. *Chemical Engineering Journal* **440** (2022). https://doi.org:10.1016/j.cej.2022.135565

(14) H.T. Yi, D. Jain, X. Yao, S. Oh. Enhanced Quantum Anomalous Hall Effect with an Active Capping Layer. *Nano Lett* **23**, 5673-5679 (2023). https://doi.org:10.1021/acs.nanolett.3c01313

(15) M.G. Shin, K.H. Bae, H.S. Jeong, et al. Effects of Capping Layers with Different Metals on Electrical Performance and Stability of p-Channel SnO Thin-Film Transistors. *Micromachines (Basel)* **11** (2020). https://doi.org:10.3390/mi11100917

(16) D.C. Vaz, E. Lesne, A. Sander, et al. Tuning Up or Down the Critical Thickness in $LaAlO_3/SrTiO_3$ through In Situ Deposition of Metal Overlayers. *Advanced Materials* **29** (2017). https://doi.org:10.1002/adma.201700486

(17) Y. Liu, S. Hu, B. Zhao, et al. Comparison of topotactic and magnetic structures for manganite oxide films. *Ceramics International* **48**, 12515-12519 (2022). https://doi.org:10.1016/j.ceramint.2022.01.118

(18) A.R. Mazza, Q. Lu, G. Hu, et al. Reversible Hydrogen-Induced Phase Transformations in $La_{0.7}Sr_{0.3}MnO_3$ Thin Films Characterized by In Situ Neutron Reflectometry. *ACS Appl Mater Interfaces* **14**, 10898-10906 (2022). https://doi.org:10.1021/acsami.1c20590

(19) E.J. Guo, Y. Liu, C. Sohn, et al. Oxygen Diode Formed in Nickelate Heterostructures by Chemical Potential Mismatch. *Advanced Materials* **30**, e1705904 (2018). https://doi.org:10.1002/adma.201705904

(20) A.K. Singh, S. Yang, U. Kar, et al. Evolution of the structural phase and orientation in ultrathin $SrRuO_3$ film with $SrTiO_3$ capping layer. *Sci Rep* **15**, 17841 (2025). https://doi.org:10.1038/s41598-025-00408-5

(21) K. Lee, B.H. Goodge, D. Li, et al. Aspects of the synthesis of thin film superconducting infinite-layer nickelates. *APL Materials* **8** (2020). https://doi.org:10.1063/5.0005103

(22) D. Li, K. Lee, B.Y. Wang, et al. Superconductivity in an infinite-layer nickelate. *Nature* **572**, 624-627 (2019). https://doi.org:10.1038/s41586-019-1496-5

(23) C.T. Parzyck, V. Anil, Y. Wu, et al. Synthesis of thin film infinite-layer nickelates by atomic hydrogen reduction: Clarifying the role of the capping layer. *APL Materials* **12** (2024). https://doi.org:10.1063/5.0197304

(24) W.J. Kim, M.A. Smeaton, C. Jia, et al. Geometric frustration of Jahn-Teller order in the infinite-layer lattice. *Nature* **615**, 237-243 (2023). https://doi.org:10.1038/s41586-022-05681-2

(25) M.I. Nandasiri, P. Nachimuthu, T. Varga, et al. Influence of growth rate on the epitaxial orientation and crystalline quality of $CeO_2$ thin films grown on $Al_2O_3(0001)$. *Journal of Applied Physics* **109** (2011). https://doi.org:10.1063/1.3525558

(26) X. Zhang, T.H. Choudhury, M. Chubarov, et al. Diffusion-Controlled Epitaxy of Large Area Coalesced $WSe_2$ Monolayers on Sapphire. *Nano Lett* **18**, 1049-1056 (2018). https://doi.org:10.1021/acs.nanolett.7b04521

(27) U. Farva, H.W. Lee, R.N. Kim, et al. Growth Temperature Influence on Atomic-Layer-Deposited $In_2O_3$ Thin Films and Their Application in Inorganic Perovskite Solar Cells. *Nanomaterials (Basel)* **11** (2021). https://doi.org:10.3390/nano11082047

(28) S. Sakthinathan, G.A. Meenakshi, S. Vinothini, et al. A Review of Thin-Film Growth, Properties, Applications, and Future Prospects. *Processes* **13** (2025). https://doi.org:10.3390/pr13020587

(29) S. Kumari, N. Mottaghi, C.Y. Huang, et al. Effects of Oxygen Modification on the Structural and Magnetic Properties of Highly Epitaxial $La_{0.7}Sr_{0.3}MnO_3$ (LSMO) thin films. *Sci Rep* **10**, 3659 (2020). https://doi.org:10.1038/s41598-020-60343-5

(30) S. Lee, X. Zhang, P. Abdollahi, et al. Route to Enhancing Remote Epitaxy of Perovskite Complex Oxide Thin Films. *ACS Nano* **18**, 31225-31233 (2024). https://doi.org:10.1021/acsnano.4c09445

(31) J.S. Lee, W.J. Kim, A. Khandelwal, et al. Cooperative Charge Ordering Signature of Trimer Molecules in Infinite-Layer $CaCoO_2$. *Physical Review X* **16** (2026). https://doi.org:10.1103/scdr-tmq4

(32) D.E. Aspnes. Spectroscopic ellipsometry—A perspective. *Journal of Vacuum Science & Technology A: Vacuum, Surfaces, and Films* **31** (2013). https://doi.org:10.1116/1.4809747

(33) A. Khare, D. Shin, T.S. Yoo, et al. Topotactic Metal-Insulator Transition in Epitaxial $SrFeO_x$ Thin Films. *Advanced Materials* **29** (2017). https://doi.org:10.1002/adma.201606566

(34) A. Korneluk, T. Stefaniuk. Multilevel Optical Storage, Dynamic Light Modulation, and Polarization Control in Filamented Memristor System. *Advanced Materials* **37**, e2411186 (2025). https://doi.org:10.1002/adma.202411186

(35) W.S. Choi, H. Jeen, J.H. Lee, et al. Reversal of the lattice structure in $SrCoO_x$ epitaxial thin films studied by real-time optical spectroscopy and first-principles calculations. *Phys Rev Lett* **111**, 097401 (2013). https://doi.org:10.1103/PhysRevLett.111.097401

(36) J.H. Gruenewald, J. Nichols, S.S. Seo. Pulsed laser deposition with simultaneous in situ real-time monitoring of optical spectroscopic ellipsometry and reflection high-energy electron diffraction. *Rev Sci Instrum* **84**, 043902 (2013). https://doi.org:10.1063/1.4798621

(37) S.S. Seo, M.J. Han, G.W. Hassink, et al. Two-dimensional confinement of $3d^1$ electrons in $LaTiO_3/LaAlO_3$ multilayers. *Phys Rev Lett* **104**, 036401 (2010). https://doi.org:10.1103/PhysRevLett.104.036401

(38) A. Seo, A. Meyer, S. Shrestha, et al. Observation of the surface layer of lithium metal using in situ spectroscopy. *Applied Physics Letters* **120** (2022). https://doi.org:10.1063/5.0096546

(39) M. Avrami. Granulation, Phase Change, and Microstructure Kinetics of Phase Change. III. *The Journal of Chemical Physics* **9**, 177-184 (1941). https://doi.org:10.1063/1.1750872

(40) W.A. Johnson. Reaction Kinetics in Processes of Nucleation and Growth. (1939).

(41) J.S. Blázquez, F.J. Romero, C.F. Conde, A. Conde. A Review of Different Models Derived from Classical Kolmogorov, Johnson and Mehl, and Avrami (KJMA) Theory to Recover Physical Meaning in Solid-State Transformations. *physica status solidi (b)* **259** (2022). https://doi.org:10.1002/pssb.202100524

(42) T. Huang, C. Zhou. Surface Diffusion-Controlled Jonhson–Mehl–Avrami–Kolmogorov Model for Hydrogenation of Mg-based Alloys. *The Journal of Physical Chemistry C* **127**, 13900-13910 (2023). https://doi.org:10.1021/acs.jpcc.3c02650

(43) K. Shirzad, C. Viney. A critical review on applications of the Avrami equation beyond materials science. *J R Soc Interface* **20**, 20230242 (2023). https://doi.org:10.1098/rsif.2023.0242

(44) S. Arrhenius. Über die Reaktionsgeschwindigkeit bei der Inversion von Rohrzucker durch Säuren. *Zeitschrift für Physikalische Chemie* **4U**, 226-248 (1889). https://doi.org:doi:10.1515/zpch-1889-0416

(45) T.K. Sandeep Kumar, N.N. Viswanathan, H. Ahmed, et al. Developing the Oxidation Kinetic Model for Magnetite Pellet. *Metallurgical and Materials Transactions B* **50**, 162-172 (2018). https://doi.org:10.1007/s11663-018-1423-4

(46) J.-i. Tani, M. Takahashi, H. Kido. Fabrication of oxidation-resistant β-$FeSi_2$ film on $Mg_2Si$ by RF magnetron-sputtering deposition. *Journal of Alloys and Compounds* **488**, 346-349 (2009). https://doi.org:10.1016/j.jallcom.2009.08.128

(47) J.J. Thomas. A New Approach to Modeling the Nucleation and Growth Kinetics of Tricalcium Silicate Hydration. *Journal of the American Ceramic Society* **90**, 3282-3288 (2007). https://doi.org:10.1111/j.1551-2916.2007.01858.x

(48) J.A. Dawson. Dynamical Insights into Oxygen Diffusion in $BaTiO_3$ and $SrTiO_3$. *physica status solidi (b)* **257** (2019). https://doi.org:10.1002/pssb.201900422

(49) R.A. De Souza, V. Metlenko, D. Park, T.E. Weirich. Behavior of oxygen vacancies in single-crystal $SrTiO_3$: Equilibrium distribution and diffusion kinetics. *Physical Review B* **85** (2012). https://doi.org:10.1103/PhysRevB.85.174109

(50) Y. Li, Y. Lei, B.G. Shen, J.R. Sun. Visible-light-accelerated oxygen vacancy migration in strontium titanate. *Sci Rep* **5**, 14576 (2015). https://doi.org:10.1038/srep14576

(51) B. Lakshminarayanan, S. Akbar, C. Wang, M. Alim. An In-house-Built Thermally Stimulated Current Measurement Setup: Strontium Titanate as a Test System.

*Japanese Journal of Applied Physics* **39**, 4830 (2000). https://doi.org:10.1143/JJAP.39.4830

(52) A. Fick. Ueber Diffusion. *Annalen der Physik* **170**, 59-86 (2006). https://doi.org:10.1002/andp.18551700105

(53) F.S. Gittleson, F. El Gabaly. Non-Faradaic $Li^+$ Migration and Chemical Coordination across Solid-State Battery Interfaces. *Nano Lett* **17**, 6974-6982 (2017). https://doi.org:10.1021/acs.nanolett.7b03498

(54) Y. Nomura, K. Yamamoto, T. Hirayama, et al. Quantitative Operando Visualization of Electrochemical Reactions and Li Ions in All-Solid-State Batteries by STEM-EELS with Hyperspectral Image Analyses. *Nano Lett* **18**, 5892-5898 (2018). https://doi.org:10.1021/acs.nanolett.8b02587